\documentstyle[12pt]{article}   
\def\be{\begin{equation}}
\def\ee{\end{equation}}
\def\beq{\begin{eqnarray}}
\def\eeq{\end{eqnarray}}
\def\n{\nonumber}
\def\bay{\begin{array}}
\def\eay{\end{array}}

\begin{document}

\begin{titlepage}

\title{Naked Singularity of the Vaidya-deSitter Spacetime and Cosmic 
Censorship Conjecture}

\author{
S.M. Wagh\dag\ddag \,\,and S.D. Maharaj\ddag
\\
\\
\dag {\footnotesize Central India Research Institute, Post Box 606,
Laxminagar, Nagpur 440 022, India} \\ 
{\footnotesize E-mail : ciri@bom2.vsnl.net.in} \\
\ddag {\footnotesize Department of Mathematics and Applied Mathematics,
University of Natal, Durban 4041, South Africa} \\ 
{\footnotesize E-mail : maharaj@scifs1.und.ac.za} \\
}

\maketitle

\begin{abstract}
We investigate the formation of a {\em locally naked singularity} in the  
collapse of radiation shells in an expanding Vaidya-deSitter
background. This is achieved by considering the behaviour of non-spacelike
and radial
geodesics originating at the singularity. A specific condition is determined
for the existence of radially outgoing, null geodesics originating
at the singularity which, when this condition is satisfied, becomes locally 
naked. This condition turns out to be the same as that in the collapse of
radiation shells in an asymptotically flat background. Therefore, we have,
at least for the case considered here, established that the
asymptotic flatness of the spacetime is not essential for the development
of a {\em locally} naked singularity. Our result then unequivocally 
supports the view that no special role be given to asymptotic observers (or,
for that matter, any set of observers) in the formulation of the Cosmic
Censorship Hypothesis. 
\end{abstract}


\noindent{\em Running head:}
Naked Singularity of Vaidya-deSitter  --- \\

\end{titlepage}

\section{Introduction}
Recently a detailed examination of several gravitational collapse scenarios
has shown [1] the development of {\em locally} naked 
singularities in a variety of cases 
such as the collapse of radiation shells, spherically symmetric self-similar
collapse of perfect fluid, collapse of spherical inhomogeneous dust cloud [2],
spherical collapse of a massless scalar field [3] and other physically
relevant situations. It is indeed remarkable that in all these cases families
of non-spacelike geodesics emerge from the naked singularity; consequently
these cases can be considered to be serious examples of 
locally naked singularity
of strong curvature type as can be verified in each individual case
separately. Such studies are expected to lead us to a proper
formulation of the Cosmic Censorship Hypothesis.

Note that all the scenarios considered so far (see [1] for details) are
spherically symmetric and asymptotically flat, and that the singularity
obtained is {\em locally naked}. We may then ask if the occurrence of
a locally naked singularity in these cases is an artefact of the special
symmetry. Or, since the real universe has no genuine asymptotically flat
objects, whether the local nakedness of the singularity in these cases is,
in some way, a manifestation of the asymptotic flatness of the solutions
considered. 

The question of special symmetry playing any crucial role in these 
situations is a hard one to settle and this possibility cannot be ruled
out easily. However, the question of asymptotic flatness playing any
special role in the development of a locally naked singularity,
at least in the collapse of radiation shells, is an easy one to settle
since the Vaidya metric in an expanding background is already known [4]. 

It is the purpose of this paper to investigate the collapse of radiation
shells in an expanding deSitter background to find out if the locally 
naked singularity occurs in this situation and to compare any difference
with the similar collapse in the asymptotically flat case. We refer the
reader to [1] for the details of the latter situation and also for 
references pertaining to it. We should point out, for the benefit of
those interested in the end result, that our conclusion is that the
locally naked singularity of the Vaidya-deSitter metric is the same as
that obtained in the asymptotically flat case. Therefore, asymptotic
flatness of the solutions considered so far does not manifest itself in
the nakedness of the singularity arising in these situations. This result
then supports the view that the asymptotic observer be not given any
special role in the formulation of the cosmic censorship hypothesis [5]
as will be discussed later.

\section{Outgoing Radial Null Geodesics of the Vaidya-deSitter Metric}
The Vaidya-deSitter metric, or the Vaidya metric in a deSitter background,
is [4]
\be ds^2\;=\;-\,\left[ 1\,-\,{{2m(v)}\over r}\,-\,\Lambda\,{{r^2}\over 3}
\right] \,dv^2\;+\;2\,dv\,dr\;+\;r^2\,d\Omega^2 \ee
where $d\Omega^2\,=\,d\theta^2\,+\,\sin^2\theta\,d\phi^2$, $v$ is the
advanced time coordinate as is appropriate for the collapse situation,
$\Lambda$ is the cosmological constant and $m(v)$ is called the mass
function. In this form the metric (1) describes the collapse of radiation.
The radiation collapses at the origin $r = 0$. 

As is well-known, the energy-momentum tensor for the radial influx of 
radiation is :
\beq T_{\alpha\beta}\;&=&\;\rho\,U_{\alpha}U_{\beta} \nonumber \\
\nonumber \\ &=&\;{1\over {4\pi r^2}}\,{{dm}\over {dv}}U_{\alpha}
U_{\beta} \eeq
where the null 4-vector $U_{\alpha}$ satisfies
$$U_{\alpha}\;=\;-\,\delta^{v}_{\alpha}, 
\;\;\;\;U_{\mu}U^{\mu} = 0 $$ 
and represents the radial inflow of radiation, in the optic limit, along the
world-lines $v = constant$. Clearly, for the weak energy condition 
$\left( T_{\alpha\beta}U^{\alpha}U^{\beta}\,\geq \,0 \right)$ we require
\be {{dm}\over {dv}}\;\geq \; 0\ee to be satisfied.

Now, let us consider the situation of radially injected flow of
radiation in an initially empty region of the deSitter universe. The
radiation is injected into the spacetime at $v = 0$ and, hence, we have
$m(v) = 0$ for $v < 0$ and the metric is that of a pure deSitter universe.
[Therefore, the inside of the radiation shells, to begin with, is 
an empty region of the deSitter metric and not the flat Minkowski metric.]
The metric for $v = 0$ to $v = T$ is the Vaidya-deSitter metric representing
a Schwarzschild field of growing mass $m(v)$ embedded in a deSitter
background. The first radiation shell collapses at $r = 0$ at time $v = 0$.
The subsequent shells collapse at $r = 0$ successivly till $v = T$ when,
finally, there is a singularity of total mass $m(T) = m_o$ at $r = 0$. 
For $v > T$, all the radiation is assumed to have collapsed and the
spacetime to have 
settled to the Schwarzschild field of constant mass $m(T) = m_o$
embedded in a deSitter background [6].

To simplify the calculations, we choose $m(v)$ as a linear function
\be 2m(v)\;=\;\lambda\,v,\hspace{0.5in}\lambda > 0 \ee
This linear mass-function was introduced by Papapetrou [7] in the 
asymptotically flat case of the Vaidya metric. Hence, in our case, the
Vaidya-Papapetrou-deSitter spacetime is described by the following
mass function for the metric (1) :
\beq m(v) = 0\hspace{.2in}
&v < 0 &\hspace{.5in}\mbox{pure deSitter} \nonumber \\
2m(v) = \lambda v \hspace{0.2in}
& 0 < v < T &\hspace{0.5in}\mbox{Vaidya-deSitter}
\\ m(v) = m_o \hspace{0.2in}& v > T &\hspace{0.5in}
\mbox{Schwarzschild-deSitter} \n
\eeq
We note at the outset that the Vaidya-deSitter spacetime for linear 
mass-function as in (5) is not homothetically Killing unlike the 
asymptotically flat Vaidya metric. In fact, the line element (1) does not
admit any proper conformal Killing symmetries.

Consider the geodesic equations of motion for the Vaidya-deSitter metric
as in (1). Let the tangent vector of a geodesic be \be K^{\alpha}\;=\;
{{dx^{\alpha}}\over {dk}}\;\equiv\;\left( \dot{v},\; \dot{r},\; \dot{\theta},
\; \dot{\phi} \right) \ee
or, equivalently,
\beq K_{\alpha} \equiv g_{\alpha\beta}K^{\beta} &=& \left( K_v,\; K_r,\; 
K_{\theta},\; K_{\phi} \right) \nonumber \\ 
&\equiv& \left( g_{vv}K^v\,+\,K^r,\; K^v,\; r^2K^{\theta},\; r^2\sin^2\theta
K^{\phi} \right) \nonumber \eeq
Then, the geodesic equations can be obtained from the Lagrangian
\be 2{\cal L}\;=\;K_{\alpha}K^{\alpha} \ee

For our purpose here it is sufficient to consider only the radiallly outgoing,
future-directed, 
null geodesics originating at the singularity. Such geodesics can be obtained
directly from the above Lagrangian as the following equation :
\be {{dv}\over {dr}} \;=\;{2\over {1 - {{2m}\over r} - {{\Lambda r^2}\over
3}}} \ee
which, for the linear mass function as in (5), is :
\be {{dv}\over {dr}} \;=\;{2\over {1 - \lambda {v\over r} - {{\Lambda r^2}
\over 3}} } \ee

Now for the geodetic tangent to be uniquely defined and to exist at the
singular point, $r = 0$, $v = 0$, of equation (9) the following must hold
\be \lim_{v \rightarrow 0\;r \rightarrow 0}\; {v\over r} \;=\;
\lim_{v \rightarrow 0\; r \rightarrow 0}\; {{dv}\over {dr}}\;=\;X_o \ee
say, and when the limit exists, $X_o$ is real and positive. In this last
situation, we obtain a future-directed, non-spacelike geodesic originating
from the singularity $r = 0$, $v = 0$ if we further demand that $2{\cal L}
\leq 0$. Then, the singularity will, at least, be {\em locally naked}. On
the other hand, if there is no real and positive $X_o$, then there is no
non-spacelike geodesic from the singularity to any observer and, hence, 
the singularity is not visible to any observer. Then, we may show that the
singularity is covered by a null hypersurface (the horizon) and the
spacetime is a black hole spacetime.

Then as we approach the singular point of the differential equation (9)
we have, using equations (9) and (10),  
\be 2\,-\,X_o\,+\,\lambda X_o^2\;=\;0 \ee
after suitable rearrangement of the terms. Thus for
the real values of the tangent to a radially outgoing, null, future-directed
geodesic originating in the singularity we obtain
\be X_o\;=\;a_{\pm}\;=\;{{1\,\pm\,\sqrt{1 - 8\lambda}}\over {2\lambda}}\ee
Clearly, we require 
\be \lambda \;\leq\;{1\over 8} \ee
for $X_o = \lim_{v \rightarrow 0\;r \rightarrow 0} v/r$ to be real in the
situation considered.

Note that the equation (10) is the same as that obtained by Dwivedi \& Joshi
[8] when the metric (1) is asymptotically flat i.\ e.\ , $\Lambda = 0$.
Consequently the values $a_{\pm}$ for the geodetic tangent and the condition
(11) for these values to be real are the same as those obtained for the
asymptotically flat situation when the mass function $m(v)$ is linear
in $v$ as in equation (4).
\goodbreak
\section{Discussion}
The present-day picture of the gravitational collapse imagines that a
sufficiently massive body compressed in too small a volume undergoes
an unavoidable collapse leading to a singularity in the very structure
of the spacetime. Of course, the deduction that a singularity will 
form as a result of such collapse tacitly assumes that we disregard
those principles of the still-ellusive quantum theory of gravity which
alter the nature of the spacetime from that given by the classical
theory of gravitation - the general theory of relativity.

Within the limits of applicability of the classical general relativity,
we characterize such unavoidable collapse by demanding the existence
of a point or of a hypersurface, called the {\em trapped surface}, whose
future lightcone begins to reconverge in every direction along the cone.
The deduction that a spacetime singularity will form is then obtained
from the well-known Hawking-Penrose Singularity Theorems [9]. These theorems 
require further physically reasonable assumptions such as the positivity
of energy and total pressure, the absence of closed timelike curves and
some notion of the genericity of the collapse situation. However, note
that the existence of a trapped surface does not imply the absence of
a naked singularity or its absence does not imply the presence of a 
naked singularity. The assumption of a trapped surface (or some other
equivalent assumption) is, however, required to infer the occurrence of
the spacetime singularity. [ See [9] for further details on this and other
related issues. ]

Now, our notion of the classical black hole situation is that of a 
spacetime singularity completely covered by an absolute event horizon. 
Unfortunately, the chronology of the  developments related
to now-famous black hole solutions emphasized the observers at future
null infinity, ${\cal I}^+$, in earlier ideas of the cosmic censor. 
We note that there is no theory concerning what happens as a result
of the appearance of a spacetime singularity. And, hence, the observer
witnessing any such singularity will not be able to account for the 
observed physical behaviour of processes involving the singularity in
any manner whatsoever. The cosmic censorship is then necessary to avoid
precisely such situations. The black hole solutions, while emphasizing
the role of observers at the future null infinity, led us into 
demanding that in the region between the absolute event horizon - the
boundary $\partial I^+ [{\cal I}^+]$ of the past of ${\cal I}^+$ - and
the set of observers at infinity, ${\cal I}^+$, no spacetime singularity
occurs.

However, it is not hard to imagine a situation in which an observer and
a collapsing body, both, are within a larger trapped surface. Thus, no
information reaches ${\cal I}^+$ from this region. But, that trapped
observer would be able to witness the forming spacetime singularity. We
are, in essence, discussing here the case of a locally naked singularity. 
For such an observer, however, it would be impossible to account for
the physical behaviour of systems involving the singularity since there
is no theory for that. The purpose of a cosmic censor, being that of 
avoiding precisely such unpredictable physical situations for legitimate
observers, is then lost on its formulation in terms of the observers
at infinity since any such formulation cannot help the above observer.

It is for avoiding such situations that we require some 
reasonable formulation of the cosmic censorship which does not single
out the set of observers at infinity. One such formulation is that of 
Strong Cosmic Censorship as given by Penrose [5].

Since our main interest here is to explore the role of asymptotic
flatness in the development of a naked singularity in the situation of
collapsing radiation shells, further analysis than that
presented in Section 2 is not necessary to draw
definite conclusions about it. The very fact that we have obtained a 
condition for the occurrence of a naked singularity in the collapse of
radiation shells in an expanding background which is the same as that 
obtained when the background is non-expanding and asymptotically flat
establishes that it is not the asymptotic flatness of the solutions
considered that manifests, in some sense, in the development of a 
{\em locally} naked singularity. In other words, whether the spacetime
is asymptotically flat or not does not make any difference to the
occurrence of a locally naked singularity. This is evident in at least
the situation of collapsing radiation shells as considered here.

Furthermore the example considered above shows that the asymptotic observer
has no role to play in the occurrence or non-occurrence of a naked 
singularity in the collapse of radiation shells. This means that the same
asymptotic observer cannot have any special role to play in the 
formulation of the Cosmic Censorship Hypothesis which is being envisaged
as a basic principle of nature, a physical law. Also, the above result
is then consistent with the viewpoint that if the cosmic censorship is to
be any basic principle of nature then it has to operate at a local level.
Hence, no special role can be given to any set of observers in the
formulation of such a basic principle; since the general theory of
relativity as a theory of gravitation provides no fundamental length
scale. Then, the present result unequivocally supports Penrose's [5] Strong
Cosmic Censorship Hypothesis which, in essence, states that singularities
should not be visible to any observer or, equivalently, no observer
sees a singularity unless and until it is actually encountered. 

\newpage

\noindent
{\Large \bf Acknowledgements :}

\bigskip
\noindent
We are grateful to Ramesh Tikekar for discussions and to an anonymous 
referee for critical reading of the manuscript and helpful suggestions.

\end{document}